\documentclass[prd,twocolumn]{revtex4}

\newcommand{\ww}{\mbox{\tiny $\wedge$}}

\begin{document}

\title{Charged solutions in 5d Chern-Simons supergravity}

\author{M\'aximo Ba\~nados}

\affiliation{Departamento de F\'{\i}sica, P. Universidad Cat\'olica de Chile, 
Casilla 306, Santiago 22,Chile.
\\ {\tt mbanados@fis.puc.cl}    }

\begin{abstract}

A family of solutions with mass and electric charge of five-dimensional Chern-Simons supergravity is displayed. The action contains an extra term that changes the value of the cosmological constant, as considered by Horava. It is shown that the solutions approach asymptotically the Reissner-Nordstrom spacetime. The role of the torsion tensor in providing charged solutions is stressed.   


\end{abstract}

\maketitle

\section{Introduction}

The Chern-Simons formulation of three-dimensional gravity \cite{Achucarro-T,Witten88} has been a useful tool to understand various issues in 3d gravity. The Brown-Henneaux \cite{BH} conformal symmetry \cite{CHvD} (see also \cite{Banados95}), the Liouville boundary action \cite{CHvD}, and the dynamical (global) degrees of freedom \cite{Witten88} have simple descriptions in the connection approach. The  black hole entropy problem also has interesting descriptions in this formulation \cite{Carlip,BBO,Mansouri-}.     

In five dimensions \cite{Chamseddine}, and actually in all odd dimensions \cite{Chamseddine,TZ,BTrZ}, there exists Chern-Simons formulations for gravity and supergravity.  These theories have a number of interesting formal properties. For example, it can be shown that the group of asymptotic symmetries, under certain specific boundary conditions, is the $WZW_4$ algebra\cite{Nair-,BGH,Losev,Gegenberg-K}. This yields  a natural extension  of the known relation between 3d Chern-Simons theory and the affine Kac-Moody algebra. Other interesting properties of this theory are the fact that the supersymmetry algebra closes off-shell, and the invariance of the action is explicit.   

The problem with higher-dimensional Chern-Simons supergravities is that they are not equivalent to standard supergravities. Since supergravity is not a fundamental theory but only an effective field theory, this should not be a reason to exclude Chern-Simons supergravity as an interesting field theory.  However, it would be desiderable to find that standard supergravity arises from Chern-Simons supergravity in some limit. For example, as a low energy approximation  on some background.              

In \cite{11d} the eleven dimensional case has been studied. It was shown that  
if the eleven-dimensional Chern-Simons supergravity action is deformed by the addition of a cosmological term, as considered by Horava \cite{Horava}, then its linear behavior around the new Anti-de Sitter background contains a three-form satisfying the correct equations.  The fermion and graviton equations expanded around this background are also the standard ones.  This result then suggests a relation between standard \cite{CJS} and Chern-Simons \cite{TZ} supergravity.  

The aim of this paper is to test other aspects of the idea introduced in \cite{11d}, in the simpler setting of five-dimensional Chern-Simons supergravity. We shall consider the $U(2,2) \sim U(1) \times SO(4,2)$ Chern-Simons action and prove that in a particular background it reproduces Einstein's theory coupled to a gauge field. In particular, the spherically symmetric configurations approach the Reissner-Nordstrom spacetime.    

The key input in this paper, as in \cite{11d}, is the introduction of a symmetry breaking term 
that changes the value of the cosmological constant.  A similar modification was consider in \cite{Horava} via the coupling of Wilson lines to the action.  Our goal is to point out that once the this term is added to the action, then its spectrum around the new anti-de Sitter background coincides with that of standard supergravity.  We shall also see that the torsion plays a key role in providing the correct quadratic kinetic term for the Abelian gauge field. 

We stress that do not claim that standard and Chern-Simons supergravity may be equivalent. Our claim is that the asymptotic behavior of Chern-Simons supergravity coincides with standard supergravity and thus it is a ``phenomenologically" interesting theory. Near the singularity  (where standard supergravity breaks down) both theories do differ.          

In Sec. \ref{ExactAction} we shall review the Chern-Simons action and its solutions with spherical symmetry. In Sec. \ref{Deformed} we introduce the deformed theory and review its uncharged solutions. In Sec. \ref{Charged} we introduce the $U(1)$ gauge field, within the Chern-Simons approach, and show that its interactions with the gravitational variables are the correct ones.     

\section{The action and equations of motion}
\label{ExactAction}

\subsection{The exact Chern-Simons action}

Let $G$ be a Lie algebra and $\Omega$ a one-form with values on $G$. The Chern-Simons action in five dimensions is defined as
\begin{equation}
I_{CS} = i \int {\cal L}_{CS} 
\label{iCS}
\end{equation}
where 
\begin{equation}
{\cal L}_{CS}= {1\over 3}\mbox{Tr} \left(d \Omega\, d\Omega\, \Omega + {3 \over 2} d\Omega \,\Omega^3 + {3 \over 5} \Omega^5 \right)
\label{CS}
\end{equation}
This Lagrangian satisfies $d{\cal L}_{CS}=(1/3)\mbox{Tr}\,(\hat R \ww \hat R \ww \hat R)$ and $\delta {\cal L}_{CS}= \mbox{Tr}\,(\hat R \ww \hat R \ww \delta\Omega)$ where $\hat R=d\Omega + \Omega\ww \Omega$.  For our choice of $\Omega$ (see appendix), the factor $i$ in (\ref{iCS}) makes the action real.

The applications of (\ref{CS}) to five dimensional supergravity follows from the observation \cite{Nahm,Chamseddine} that the superalgebra $SU(2,2|N)$ contains the anti-de Sitter group.   
For this choice of the Lie algebra, the action (\ref{CS}) yields a Lovelock type theory of gravity coupled to fermions and some gauge fields. The whole action is invariant under adjoint transformations acting on $\Omega$ and is thus supersymmetric.  

We shall be interested in the bosonic degrees of freedom, in particular in the corresponding action associated to the Lie algebra $U(2,2)$ containing the gravitational variables plus an Abelian gauge field.  The Lagrangian expressed in terms of the $e^a,w^{ab}$ and the Abelian field $A$ is \cite{Chamseddine} (see the Appendix A for a detailed derivation), 
\begin{equation}
i{\cal L}_{CS}  = i{\cal L}_{G}  + i{\cal L}_A + i{\cal L}_I
\label{U22}
\end{equation} 
where
\begin{eqnarray}
i{\cal L}_G &=& -{1 \over 8}\epsilon_{abcde} \left({1 \over l} R^{ab}\ww R^{cd}\ww
e^e +  {2 \over 3l^3 } R^{ab}\ww e^c \ww e^d \ww e^e  \right. \nonumber  \\ 
     & & \ \ \ \ \ \ \ \ \ \ \  \ \ \ \    \left.  + {1 \over 5 l^5 } e^a\ww e^b\ww e^c\ww e^d\ww e^e  \right) \label{CSG} \\
i{\cal L}_A  &=&  - {\chi \over 3} \, A\ww  F \ww F  \\
i{\cal L}_I  &=& {1 \over 8} R^{ab}\ww R_{ab}\ww A - {1 \over 4l^2}  e_a T^a \ww F.
\end{eqnarray}
and $\chi = N^{-2} - 4^{-2}$.

${\cal L}_G$ is the purely gravitational Lagrangian. This term is a Chern-Simons Lagrangian in its own right for the group $SO(4,2)$.  Note that both the Hilbert and cosmological terms are present. ${\cal L}_G$ can also be regarded as a Lovelock Lagrangian with a prescribed choice of coefficients. This choice of coefficients have two important consequences: (i) they enlarge the local symmetry from Lorentz $SO(4,1)$ to $SO(4,2)$, (ii) the equations of motion have the structure $ (R + \Lambda )^2=0$ making the anti-de Sitter background degenerate.

${\cal L}_A$ is an Abelian CS Lagrangian which is also present in the standard five-dimensional supergravity.  

${\cal L}_I$ is the interaction term between $A$ and the gravitational variables. Up to a boundary term, this piece can also can also be written as $(\bar R^{ab} \bar R_{ab} - (2/l^2) T^a T_a)A/4$.  
    
Varying the Lagrangian (\ref{U22}) with respect to $e^a$, $w^{ab}$ and $A$ one obtains the equations of motion, 
\begin{eqnarray}
\epsilon_{abcde} \bar R^{ab} \mbox{\tiny $\wedge$} \bar R^{cd}  &=&   -{4\over l}\,  T_e
\mbox{\tiny $\wedge$} F,
\label{e} \\
{1 \over l}\epsilon_{abcde} \bar R^{ab} \mbox{\tiny $\wedge$} T^c  &=& \bar R_{de} \mbox{\tiny
$\wedge$} F, 
\label{w} \\
{1 \over 2} R^{ab} \mbox{\tiny $\wedge$} R_{ab} -{1 \over l^2} d(e_a \mbox{\tiny
$\wedge$} T^a)  &=& 4\chi F \mbox{\tiny $\wedge$} F,  \label{A} 
\end{eqnarray}
respectively. $T^a = De^a$ is the torsion tensor, $F=dA$ and 
\begin{equation}
\bar R^{ab} = R^{ab} + {1 \over l^2} \, e^a \mbox{\tiny $\wedge$} e^b
\label{Rbar}
\end{equation}
with $R^{ab}=dw^{ab} + w^a_{\ c}w^{cb}$.  Note that all coefficients in
(\ref{e},\ref{w},\ref{A}) are fixed (up to trivial rescallings) by supersymmetry. Equations (\ref{e},\ref{w},\ref{A}) have the generic form
\begin{equation}
g_{abc} F^b \ww F^c =0
\end{equation}
where $F^a\in U(2,2)$ and $g_{abc}$ is an invariant tensor. 

\subsection{Chern-Simons black holes}

In the uncharged case, $F=0$, black holes solutions associated to (\ref{e},\ref{w},\ref{A}) have been  discussed in \cite{BTZ2}.  Let us briefly review and extend to the case $F\neq 0$ \footnote{Charge was introduced in \cite{BTZ2} by adding to the action the term $\sqrt{-g}\, F^{\mu\nu}F_{\mu\nu}$, thus breaking the Chern-Simons structure.} the solutions found in \cite{BTZ2}. This is instructive because it gives the first hint that the exact Chern-Simons theory described by (\ref{CS}) does not reproduce Einstein's gravity.   

Consider the spherically symmetric ansatz 
\begin{eqnarray}
ds^2 &=& - f(r)^2 dt^2 + {h(r)^2 \over f^{2}(r)} dr^2 + r^2 d\Omega_3  \label{g0}\\
  A  &=& -\phi(r) dt,
\end{eqnarray}
and zero torsion. Replacing in (\ref{e},\ref{w},\ref{A}) one finds the solution for the functions $f,h$ and $\phi$, 
\begin{eqnarray}
   f(r)^2 &=& {r^2\over l^2} + f_0, \label{f0} \\
   h(r) &=& 1,\\
\phi(r) &=& \phi_0,
\label{exact}
\end{eqnarray}
where $f_0$ and $\phi_0$ are arbitrary constants.

From this solution we learn that the spherically symmetric Abelian field does not couple to the gravitational variables, and the metric is not asymptotically Schwarzschild-AdS. The only allowed perturbation is a constant in the function $f(r)^2$, which is not interesting from a  ``phenomenological" point of view. 

Near the singularity, this solution takes the form, 
\begin{equation}
ds^2 = - f_0\, dt^2 + {1 \over f_0} dr^2 + r^2 d\Omega_3.
\label{cone}
\end{equation}
If $f_0=1$ this is regular. For other values of $f_0$ one finds ``conical" singularities in the plane $r,\varphi$.  If $f_0<0$ then the singularity lies behind the horizon.  Below, we shall modify the Chern-Simons action in order to produce asymptotically Schwarzschild spacetimes. However, the solution (\ref{g0}) will reappear in the near singularity (and horizon) region. 

Equations (\ref{e}-\ref{A}) differ drastically from the standard Einstein-Electromagnetic system.  Consider, for example the Abelian field $A$ propagating on the gravitational AdS background with $\bar R^{ab}=T^a=0$. The gravitational equations (\ref{e}) and (\ref{w}) are identically satisfied while the gauge field equation yield 
\begin{equation}
\chi F \mbox{\tiny $\wedge$} F=0,
\end{equation}
which is quite different from the expected equation  
\begin{equation}
d^*\! F =  F \ww F
\label{fieldE}
\end{equation}

\section{The deformed Chern-Simons theory} 
\label{Deformed}

\subsection{The action and equations of motion }

As shown in the last section, the exact Chern-Simons theory does not have asymptotically Schwarzschild solutions, and the gauge field $A$ does not satisfy its standard equations of motion.  These two problems can be solved at once by breaking the symmetry of the Chern-Simons action.   

Suppose we add to (\ref{CS}) a contribution to the cosmological term, 
\begin{equation}
{\cal L}_\tau = {\cal L}_{CS} + {\tau^{2} \over 40l^5} \epsilon_{abcde}e^a \ww \cdots \ww e^e
\label{Itau}
\end{equation}
where $\tau$ is a dimensionless number parametrizing the deformation. An interesting mechanism to generate this term has been proposed in \cite{Horava}, via the coupling of Wilson lines. For our purposes it will not be necessary to fix $\tau$ to any particular value, although some constraints will be found below. 

The new equations of motion are 
\begin{eqnarray}
\epsilon_{abcde} (\bar R^{ab} \mbox{\tiny $\wedge$} \bar R^{cd}- {\tau^2\over l^4}  e^a\ww e^b\ww e^c \ww e^d )  &=& - {4\over l} \,  T_e \mbox{\tiny $\wedge$} F   
\label{edef} \\
{1 \over l}\epsilon_{abcde} \bar R^{ab} \mbox{\tiny $\wedge$} T^c &=& \bar R_{de} \mbox{\tiny
$\wedge$} F  
\label{wdef} \\
{1 \over 2} R^{ab} \mbox{\tiny $\wedge$} R_{ab} - {1 \over l^2}d(e_a \mbox{\tiny
$\wedge$} T^a) &=& \chi F \mbox{\tiny $\wedge$} F   
\label{Adef}
\end{eqnarray}

\subsection{Torsion and a dynamical gauge field}

Let us first show that the new equations yield the desired equation (\ref{fieldE}) for the gauge field \footnote{See \cite{dAuria} for a description of this mechanism within the group manifold approach to supergravity, and \cite{Sardinia} for a generalization.}. We freeze the gravitational degrees of freedom ($e^a$) and study the dynamics of $A$. We shall also discard all quadratic terms in $F$ and $T^a$ keeping only the linear terms.

Due to the new term in (\ref{edef}) the correct gravitational background is not $\bar R^{ab}=0$ but 
\begin{equation}
\bar R^{ab} = {\tau \over l^2} e^a \ww e^b,
\label{tauAdS}
\end{equation} 
corresponding to a constant curvature spacetime with 
\begin{equation}
\Lambda_\tau = - {1 -\tau \over l^2}.
\label{L}
\end{equation}
Note that for $\tau >0$ big enough $\Lambda$ becomes positive and the background is de Sitter.

We now replace  (\ref{tauAdS}) in (\ref{wdef}) and obtain,
\begin{equation}
{\tau\over l} \epsilon_{abcde} e^a\ww e^b \mbox{\tiny $\wedge$} T^c =\tau e_d\ww e_e \mbox{\tiny
$\wedge$} F  \ \ \ \Rightarrow \ \ \ {2\over l} e_a \ww T^a = \, ^* \! F
\label{eT=F}
\end{equation}
provided $\tau \neq 0$.     
The second equality follows straightforwardly from the first one. Note, in particular, that  $T_{\mu\ \nu\rho}=e_{a\mu}\, T^a_{\ \nu\rho}$ is totally antisymmetric. Finally, replacing  (\ref{eT=F}) in (\ref{Adef}) we obtain the desired (linear) equation,
\begin{equation}
d^*\! F = 0
\label{A3} 
\end{equation}
for an Abelian gauge field \footnote{These is actually a contribution to this equation from $R^{ab}R_{ab}$ as well. On the AdS background $R^{ab} = -\Lambda_\tau e^a\ww e^b + D \kappa^{ab}$. Using (\ref{eT=F}) we find that $\kappa_{ab} e^a e^b =\, ^*\!F$ and then $R^{ab}\ww R_{ab} = \Lambda d^*\! F$.}. 

The main lesson of this result is that the spin connection contains both the second order structure for the graviton and the second order structure for the gauge field. In fact, the spin connection has the form
\begin{equation}
w^{ab} = w^{ab}(e) + \kappa^{ab}(A)
\label{wkappa}
\end{equation}
where $w^{ab}(e)$ is the torsion-free part, while $\kappa(A)$ is the torsion and depends on $A$ as in (\ref{eT=F}). 

When replacing (\ref{wkappa}) back into the action, the term $w^{ab}(e)$ produces the standard graviton second order kinetic term $\int e R(e)$, while $\kappa^{ab}(A)$ produces the term  $\int e F^{\mu\nu}F_{\mu\nu}$.  Interestingly, in eleven dimensions, the supergravity three-form \cite{CJS} arises via a similar mechanism from the torsion \cite{11d} in a deformed Chern-Simons theory.  This mechanism has also appeared in the Group Manifold approach to supergravity \cite{dAuria}.

\subsection{Uncharged exact solution}

Let us now set $F=T^a=0$ and study the gravitational solutions of the deformed theory. The solution to (\ref{edef}-\ref{Adef}) with spherical symmetry is
\begin{equation}
ds^2 = - f(r)^2 dt^2 + {1\over f(r)^2} dr^2 + r^2 d\Omega_3  
\end{equation} 
with \cite{Boulware-D,Wheeler}
\begin{equation}
f(r)^2 = 1 + {r^2 \over l^2} - \tau \, \sqrt{ {r^4\over l^4} + {C_0\over l^2 \, \tau^2}}
\label{Boulware-D}
\end{equation}
and $C_0$ is an arbitrary integration constant proportional to the mass. In the limit $\tau\rightarrow 0$ we recover the solution of the unbroken theory (\ref{f0}), and for $C_0=0$ we recover the background (\ref{tauAdS}) of the deformed theory.  In what follows we shall only consider the case 
\begin{equation}
\tau >0
\end{equation}
since otherwise $f(r)^2$ has no real zero'es and no horizon.         

The interesting property of this solution is that for large values of $r$, the function $f(r)^2$ takes the form
\begin{equation}
f(r)^2 =  -\Lambda_\tau \, r^2 + 1 - {2m \over  r^2}+ {\cal O}\left( {1 \over r^6} \right) \ \ \ \  (r\rightarrow \infty)
\end{equation} 
coinciding, as promised, with the asymptotic form of the Schwarschild-AdS spacetime in five dimensions ($m= C_0/(4\tau)$ and $\Lambda$ is given in (\ref{L})). It is interesting to note that the term $1/r^4$ is absent in the asymptotic expansion of $f(r)^2$. As expected, this term will arise in the charged case.     

On the other hand, near the singularity ($r\rightarrow 0$) the deformation parameter $\tau$ becomes unimportant, 
\begin{equation}
f(r)^2 = 1 -  {C_0^{1/2} \over l},  \ \ \ \  (r\rightarrow 0)
\end{equation}
and the solution approaches (\ref{cone}).   The horizon structure is also controlled by the exact Chern-Simons theory. In the AdS sector with $\Lambda_\tau<0$ ($1-\tau>0$), the equation $f(r_+)^2=0$ has one zero \footnote{If $1-\tau <0$ there are two zero's and the solution is de Sitter}, 
\begin{equation}
r_+^{\ 2} = {\sqrt{C_0(1-\tau^2) - \tau^2}-1  \over 1-\tau^2 },
\label{r+}
\end{equation}
which is real provided $C_0>1$. The spectrum of the solution then coincides with the Chern-Simons black hole \cite{BTZ,BTZ2} 

~

\centerline{
\begin{tabular}{rl}
$C_0 = 0:$ \ \   & \ \ Background (regular)\\
$0<C_0 < 1:$ \ \  &  \ \ Naked (``conical") singularities  \\
$C_0 \geq 1:$ \ \ & \ \ Black holes
\end{tabular}}

~
  
In summary, the deformation parameter $\tau$ is important in the asymptotic region producing ``phenomenologically" interesting solutions. Near the singularity the physics is controlled by the exact Chern-Simons theory. The function $f$ is regular at $r=0$ but the geometry is not. For example the scalar curvature diverges as \cite{BTZ2}, 
\begin{equation}
R \sim {C_0 \over l^2 r^2}.  
\label{Rdiv}
\end{equation}
It is worth noticing that this divergence is milder than that of Schwarzschild $R\sim m /r^4$.

\section{The charged solution} 
\label{Charged}

In this section we study charged ($F\neq 0$) solutions of the deformed Chern-Simons theory described by the equations (\ref{edef}-\ref{Adef}).  

We shall see that the interaction between the gauge field and gravitational variables is,  asymptotically, the usual minimal coupling.  Equations (\ref{edef},\ref{wdef},\ref{Adef}) have  the asymptotic Reissner-Nordstrom solution, 
\begin{eqnarray} 
ds^2 &=& - f(r)^2 dt^2 + {1 \over f(r)^2} dr^2  + r^2 d\Omega_3, \label{metric2} \\
   A &=& -\phi(r) dt,
\end{eqnarray}
where the leading terms of $f$ and $\phi$ are 
\begin{eqnarray}
f(r) &=& -\Lambda r^2  + 1 - {C_0 \over r^2} + k \, {q^2 \over r^4} \label{f2} \\ 
\phi(r) &=& { q \over r^2}.
\end{eqnarray}
$\Lambda$ is given in (\ref{L}). The coupling constant $k$ turns out to be 
\begin{equation}
k={l^2 \over 20} \, \left(29 - {24 \over \tau}\right).
\label{k}
\end{equation}
In particular, we find the condition $\tau  > 24/29$ in order to have the correct sign.

The full set of equations when $F\neq 0$ is an extremely complicated non-linear system. We have solved these equations perturbatively starting from the vacuum (\ref{tauAdS}), and this yields  the functions $f$ and $\phi$ displayed above. The goal of this section is to expand up to second order and prove the appearance of the term $k\, q^2/r^4$ in (\ref{f2}), showing that the interaction terms between gravity and the gauge field have the desired asymptotic behavior.

\subsection{The ansatz and solution}

As we have seen, the correct dynamical theory for $A$ is linked to the torsion tensor. As usual when dealing with spacetimes with torsion we define 
\begin{equation}
w^{ab} = w^{ab}(e) + \kappa^{ab}
\label{kappa}
\end{equation} 
where $w^{ab}(e)$ is the solution to the equation $de^a + w^a_{\ b}(e) \ww e^b=0$, and $\kappa$ is related to the torsion as $T^a=\kappa^a_{\ b} \ww e^b$. Assuming that $e^a_\mu$ is invertible, the relation between $T^a_{\ \mu\nu}$ and $\kappa^{ab}_{\ \ \mu}$ is invertible. 

The variables of this problem are then $e^a$, $\kappa^{ab}$ and $A$.  We start by assuming the spherically symmetric ansatz for the gauge field and metric
\begin{eqnarray} 
ds^2 &=& - f(r)^2 dt^2 + {h(r)^2 \over f(r)^2} dr^2  + r^2 d\Omega_3, \label{gsph}\\
   A &=& -\phi(r) dt,  \label{Asph}
\end{eqnarray}
where $f,h$ and $\phi$ are functions to be determined.  

The situation is more complicated for the torsion,  
\begin{equation}
\kappa_{\alpha\beta\ \mu } =  e_{a\alpha}\, \kappa^{ab}_{\ \ \mu} \, e_{b\beta},
\label{kappamu}
\end{equation}
because this tensor has no prescribed symmetries between $\{\alpha,\beta\}$ and $\mu$. The problem is then to determine which components in the irreducible decomposition of $\kappa$ will contribute to the spherically symmetric solution.  We have already seen in Eq. (\ref{eT=F}) that to first order, only the purely antisymmetric part of $\kappa$ contributes.  This is however no longer true in the full solution. 

In the Appendix \ref{kappaAppen} we display a systematic procedure to find $\kappa$ for the spherically symmetric solution.  This ansatz has the form, 
\begin{equation}
\kappa_{\alpha\beta\ \mu } = \theta_{\alpha\beta\mu}  + z_{\alpha\beta}\, 
U_\mu + g_{\mu [\alpha } V_{\beta]}  
\label{kappafull}
\end{equation}
where $\theta$ and $z$ are both fully antisymmetric in their indices,  
\begin{eqnarray}
^* \! \theta &=& \psi(r) \, dt \ww dr,  \label{U} \\
z       &=& dr \ww dt, \\
U     &=& \beta(r) \, dt, \\
V       &=& \alpha(r) \, dr, \label{C}
\end{eqnarray}
where $*$ denotes Hodge's dual (in five dimensions, the dual of $\theta$ is a two form). $\psi,\beta,\alpha$ are unknown functions of $r$ to be determined.   

The full ansatz then contains 6 functions of the radial coordinate, namely, $f$ and $h$ appearing in the metric (\ref{gsph}), $\phi$ is the Coulomb potential in (\ref{Asph}), and $\psi,\alpha,\beta$ appear in the torsion. We now plug this ansatz in the equations (\ref{edef}-\ref{Adef}) and find, unfortunately, a complicated non-linear system of equations which cannot be solved in a closed form.  What we can do is to analyze it perturbatively starting from the known background (\ref{tauAdS}).          

To this end, we expand each function in the ansatz up to second order, 
\begin{eqnarray}
f{\,^2} &=& \stackrel{(0)}{u} + \sigma \stackrel{(1)}{u} + \sigma^2 \stackrel{(2)}{u}, \\
h &=& 1 + \sigma \stackrel{(1)}{h} + \sigma^2 \stackrel{(2)}{h}, \\
\phi &=&  \sigma \stackrel{(1)}{\phi} + \sigma^2 \stackrel{(2)}{\phi}, \\
\psi &=& \sigma \stackrel{(1)}{\psi} + \sigma^2 \stackrel{(2)}{\psi}, \\ 
\alpha &=&  \sigma \stackrel{(1)}{\alpha} + \sigma^2 \stackrel{(2)}{\alpha}, \\
\beta &=& \sigma \stackrel{(1)}{\beta} + \sigma^2 \stackrel{(2)}{\beta}, 
\end{eqnarray}
where $\sigma$ is the expansion parameter which will be set equal to one at the end ($\sigma$ can be absorbed in the charges that will appear in the solution). At order zero, we set  the anti-de Sitter background (\ref{tauAdS}) with 
\begin{equation}
\stackrel{(o)}{u} = -\Lambda_\tau\, r^2 + 1.
\end{equation}
($\Lambda_\tau$ is given in (\ref{L}).)  

We shall skip the detailed calculation of the equations order by order because it is rather long and devoid of any interesting physics.  Instead we present a summary of the results.    

At order one, we find the five-dimensional Newton and Coulomb potentials 
\begin{equation}
\stackrel{(1)}{u} = -{C_0 \over r^2}, \ \ \ \  \stackrel{(1)}{\phi} = {q \over r^2}, 
\end{equation}
where $C_0$ and $q$ are, respectively, proportional to the mass and electric charge. At this order only the fully antisymmetric part of torsion is different from zero, 
\begin{equation}
 \stackrel{(1)}{\psi} \! (r) = {q \over 2r^3}, \ \ \ \ \ 
\stackrel{(1)}{\alpha} =  \stackrel{(1)}{\beta}=\stackrel{(1)}{h}=0.  
\end{equation}

At order two we obtain the back reaction from the gauge field to the metric
\begin{equation}
\stackrel{(2)}{u} = k'\, {q^2 \over r^4} 
\end{equation}
having the expected form, with $k'=l^2(19/12 - 4/(3\tau)) $. 
At this order, the torsion is not fully antisymmetric, and there are also further corrections to $h$ and $\phi$,
\begin{eqnarray}
\stackrel{(2)}{h}    &=&  {l^4 q^2 \over 3 r^6 \, \tau }, \\
\stackrel{(2)}{\phi} &=& -{l^2 q C_0  \over 3 r^6\, \tau}, \\
\stackrel{(2)}{\psi} &=& {l^3 q C_0  \over 2 r^7 \, \tau  }, \\
\stackrel{(2)}{\alpha} &=& {l^4 q^2 \over r^7 \, \tau}, \\
\stackrel{(2)}{\beta}  &=& -{4 l^4 q^2 \over r^7\, \tau } \, (1 -\Lambda r^2).     
\end{eqnarray}

Note that at second order, the function $h$ is not equal to one, and thus the form of the metric (\ref{gsph}) is not the standard one.  One can me make a radial redefinition 
$h(r)dr \rightarrow dr $ which leads to the metric (\ref{metric2}).  (The correction in $r^2 d\Omega_3$ is subleading.)

The particular form of the coupling (\ref{k}) can be understood from the Chern-Simons Lagrangian (\ref{U22}). We freeze the gravitational variable $e^a$, and replace the torsion in terms of $F$ as in Eq. (\ref{eT=F}). Expanding to second order in $F$, one finds various contributions to $^*\! F \,F$.  For example the term $e_a T^a F$ yields $^* \!F\, F$, while $R^{ab} R_{ab} A$ yields $\Lambda_\tau \, ^* \!F \, F$. The sum of all terms produce the above coupling.

\section{Conclusions}

We have shown in this paper that if the five dimensional \cite{Chamseddine} Chern-Simons supergravity action is deformed by the addition of cosmological term, then its asymptotic behavior coincides with that of standard supergravity. 

The role of torsion in providing charged solutions is particularly interesting. It has been shown in \cite{11d} that a similar mechanism holds in eleven dimensions. In this case, the three-form \cite{Nahm,CJS} arises as the fully antisymmetric part of the torsion.  

It would be interesting to find an exact solution to the equations in the charged case. This would allow us to characterize the charged black hole solutions in Chern-Simons theory.  In particular, it would be interesting to check whether the deformation $\tau$ is unimportant near the origin or not (as in the uncharged case), and to study the singularity structure and spectrum of the charged black hole.  The existence of extreme black holes and their supersymmetric properties would be particularly interesting.

\appendix

\section{The five-dimensional U(2,2) Chern-Simons Action} 

The goal of this appendix is to give some details in the derivation of the Lagrangian (\ref{U22}) starting from (\ref{CS}). This has been discussed in \cite{Chamseddine,TZ},  however, there appear to be some misprints in those References. Since in our applications (see also \cite{selfdual}) all terms and numerical factors are important, we include here the derivation. 

We use the Dirac matrices in five dimensions 
\begin{equation}
\{\gamma_a,\gamma_b\}=-2\eta_{ab}
\label{Dirac}
\end{equation}
with $\eta_{ab} = \mbox{diag}(-1,1,1,1,1)$. The following traces will be needed, 
\begin{eqnarray}
\mbox{Tr} (\gamma_a\, \gamma_b) &=& -4 \eta_{ab}, \\
\mbox{Tr} (\gamma^{ab}\, \gamma_{cd}) &=& - 4\delta^{[ab]}_{\, [cd]} \\
\mbox{Tr} (\gamma_a\, \gamma_b\, \gamma_c \, \gamma_d \, \gamma_e) &=& 4\epsilon_{abcde}  
\label{traces}
\end{eqnarray} 
with $\epsilon_{01235}=1$ and $\gamma_5=\gamma_0\gamma_1\gamma_2\gamma_3$. Note that $\gamma_0^\dagger = \gamma_0$ while $\gamma_i^\dagger = - \gamma_i$ for $i=1,2,3,5$. $\gamma_0$ is taken to be 
\begin{equation}
\gamma_0 =\left( \begin{array}{cccc}   1 &  &   &     \\
                                       & 1 &  &   \\  
                                       &   &-1  &\\
                                       &   &    & -1\\  
                                      \end{array} \right).
\end{equation}

\subsection{The U(2,2) gauge field }

The first step is to express the $SU(2,2|N)$ gauge field $\Omega$ in terms of the veilbein, spin connection and Abelian gauge field $A$. By definition, $\Omega$ is a $(N+4)\times (N+4)$ complex matrix satisfying
\begin{equation}
(\Omega \Gamma )^\dagger = - \Omega \Gamma, \ \ \ \ \ \
\mbox{STr} (\Omega)=0,
\label{su}
\end{equation}
where the $(N+4)\times (N+4)$ matrix $\Gamma$ is 
\begin{equation}
\Gamma = \left( \begin{array}{cc}   \gamma_0 &  0  \\
                           0 &  \mbox{I}_N   \end{array} \right)
\label{Gamma}
\end{equation}
and I$_N$ is the identity in $N$ dimensions. 

Conditions (\ref{su}) can be implemented by expanding $\Omega$ in the form \cite{Chamseddine}
\begin{equation}
\Omega = \left( \begin{array}{cc}   W &   \psi   \\
                          -\bar\psi &  {\cal A}   \end{array} \right)
\label{Omega}
\end{equation}
where $W\in U(2,2)$ i.e., $(W\gamma_0)^\dagger = -W\gamma_0$, $\psi$ is a Dirac
spinor in five dimensions ($\bar \psi = \psi^\dagger \gamma_0$) and ${\cal A}
^\dagger = - {\cal A}$ is a $N\times N$ matrix.  The super trace of $\Omega$ is required to vanish and hence Tr$\, W - \mbox{Tr} {\cal A}=0$, but the individual traces of $W$ and ${\cal A}$ can be non-zero.  

In what follows we set the spinor fields to zero. We shall also discard the $SU(N)$ fields (traceless part of ${\cal A}$). For the full Lagrangian see \cite{Chamseddine,TZ}. 

A basis of 4$\times$ 4 matrices satisfying $(M\gamma_0) ^\dagger = -M\gamma_0$ is given by the matrices $i\gamma_a,\gamma_{ab}$ and $i\, I_4$. We then expand
\begin{eqnarray}
W &=& {i \over 4} A \ \mbox{I}_4 + {i \over 2l} e^a \gamma_a - {1 \over 4}
w^{ab} \gamma_{ab}, \label{W} \\
{\cal A} &=& {i \over N} A\ \mbox{I}_N.
\end{eqnarray}
where $A$ is an Abelian gauge field. Note that Tr$\, W = \mbox{Tr} {\cal A}$ as desired. $e^a$ and $w^{ab}$ are identified with the vielbein and spin connection, respectively.  It will also be convenient to  define the $SU(2,2)\sim SO(4,2)$ field, 
\begin{equation}
W' = {i \over 2l} e^a \gamma_a - {1 \over 4} w^{ab} \gamma_{ab}
\label{W'}
\end{equation}
carrying the gravitational variables.

\subsection{The Lagrangian}

Setting the spinors to zero, the gauge field (\ref{Omega}) has a block diagonal form and replacing in (\ref{CS}) we obtain, 
\begin{equation}
{\cal L} = {\cal L}(W) +  {i \over 3 N^2} FFA 
\end{equation}
where 
\begin{equation}
{\cal L}(W) = {1 \over 3}\mbox{Tr} ( dW dW W + {3 \over 2} dW W^3 + {3 \over 5} W^5)
\end{equation}
The $U(2,2)$ field $W$ contains the gravitational variables as well as the Abelian gauge field, 
\begin{equation}
W = W' + {i \over 4} A \, \mbox{I}_4
\end{equation}
where $W'$ is given in (\ref{W'}).  There is a simple way to write ${\cal  L}(W)$ as a function of $W'$ and $A$.  We shall prove the following formula
\begin{equation}
{\cal L}(W) = {\cal  L}(W')  -{i \over 3 \cdot 4^2} FFA + {i \over 4}\mbox{Tr} (R'R') A
\label{LW}
\end{equation}
where $R'= dW' + W'W'$.  The first two terms in (\ref{LW}) are obvious because ${\cal L}$ is a polynomial in $W$ which must reduce to ${\cal  L}(W')$  if $A=0$, and to ${\cal  L}(A)$ if $W'=0$.  The interaction term can be determined as follows.  Since $W' \in SU(2,2)$ with $\mbox{Tr}R'= 0$ and $A$ is Abelian, the curvature $R'$ can only appear quadratically in the interaction term. This implies that $A$ must appear linearly and therefore it can be treated as infinitesimal. The well known formula for the variation of the Chern-Simons action, 
\begin{equation}
{\cal L}(W'+ X) = {\cal L}(W') +  \mbox{Tr} (R' R' X)
\end{equation}         
can then be used and leads to the third term in (\ref{LW}) with $X=iA/4$.  Next we note that $R'=dW'+W'W'=(i/2) T^a\gamma_a - (1/4) \bar R^{ab} \gamma_{ab}$ and hence 

\begin{eqnarray}
\mbox{Tr}\, (R'R') &=& - {1 \over 2} \bar R^{ab}\ww \bar R_{ab} +  T^a\ww T_a. 
   \label{R'R'}
\end{eqnarray}
 
Finally, we need to write ${\cal L}(W')$ in terms of $e^a$ and $w^{ab}$. Up to boundary terms one has the equality \cite{Chamseddine,BTZ2}
\begin{eqnarray}
{1 \over 3}\mbox{Tr} ( dW' dW' W' + {3 \over 2} dW' W^{'3} + {3 \over 5} W^{'5})                 &=&\nonumber \\  k_0 \, \epsilon_{abcde} \left({1 \over l} R^{ab}\ww R^{cd}\ww
e^e +  {2 \over 3l^3 } R^{ab}\ww e^c \ww e^d \ww e^e + \right. \nonumber \\ \left. {1 \over 5 l^5 } e^a\ww e^b\ww e^c\ww e^d\ww e^e  \right)      
\end{eqnarray}
where $W'$ is related to $w^{ab}$ and $e^a$ in (\ref{W'}). 
The constant $k_0$ can be fixed by choosing $w^{ab}=de^a =0$. Comparing both sides one finds $k_0 = i/8$.  

Collecting all terms one arrives at the final formula displayed in  (\ref{U22}).

\section{The ansatz for $\kappa$}
\label{kappaAppen}

The correct form for $\kappa^{ab}$ in the spherically symmetric ansatz can be obtained via a systematic expansion around the deformed background (\ref{tauAdS}). We consider the expansion,
\begin{eqnarray}
e^a  &=&  \stackrel{(o)}{e} \! ^a + \sigma \stackrel{(1)}{e} \! ^a + \sigma^2 \stackrel{(2)}{e}\! ^a \\
A  &=&  \sigma \stackrel{(1)}{A} + \sigma^2\stackrel{(2)}{A} \\ 
\kappa^{ab}  &=& \sigma \stackrel{(1)}{\kappa} \! ^{ab} + \sigma^2 \stackrel{(2)}{\kappa}\! ^{ab} 
\end{eqnarray}
At order zero we have the background (\ref{tauAdS}),
\begin{eqnarray}
\stackrel{(o)}{\bar R}\! ^{ab} = {\tau \over l^2}  \stackrel{(o)}{e}\! ^a \ww  \stackrel{(o)}{e}\! ^b.
\end{eqnarray}
The torsion is zero at this order, hence the curvature depends only on the vielbein $\stackrel{(o)}{e} \! ^a$.  By looking at the equations of motion order by order we will be able to find the form of $\kappa$ at each order.   

At order one, the torsion $\kappa$ and electromagnetic field $A$ are coupled through (\ref{wdef}). We already encountered this equation in (\ref{eT=F}). In particular, this means that $\stackrel{(1)}{\kappa}\! ^{ab}_{\ \ \mu}$ is fully antisymmetric, 
\begin{equation}
\stackrel{(1)}{\kappa}\!_{\mu\nu\rho} = \epsilon_{\mu\nu\rho\alpha\beta} \stackrel{(1)}{F}\! ^{\alpha\beta}.
\end{equation}
Thus, given the Coulomb ansatz for $ \stackrel{(1)}{A}$, the first order function  $\stackrel{(1)}{\kappa}$ is completely determined.  

At order two, the equation for the torsion (\ref{wdef}) gives, 
\begin{eqnarray}
\tau \epsilon_{abcdf} \stackrel{(o)}{e}\!^a \ww \stackrel{(o)}{e}\!^b \ww \stackrel{(2)}{T}\! ^c &=& \tau \stackrel{(o)}{e}\!_d \ww \stackrel{(o)}{e}\!_f \ww \stackrel{(2)}{F} +  \\ && \stackrel{(1)}{\bar R} \! _{df} \ww \stackrel{(1)}{F} - \epsilon_{abcdf} \stackrel{(1)}{\bar R} \! ^{ab} \ww \stackrel{(1)}{T} \! ^c   \nonumber
\end{eqnarray}
From this equation we can solve  $\stackrel{(2)}{T}\! ^a$ in terms of  $\stackrel{(2)}{F}$ and the first order fields.  Since the spherically symmetric form of $F$ is known at all orders  from this equation we can find the non-zero components of $\stackrel{(2)}{\kappa}$. This calculation is straightforward and we shall not present the details.  One finds that $\stackrel{(2)}{\kappa} $ contains a trace, and a mixed component. The explicit form for the full $\kappa$ is given in (\ref{kappafull}).

\acknowledgments

The author would like to thank A. Gomberoff for useful discussions. This work was partially supported by grant \# 1000744 (FONDECYT, Chile).


\begin{thebibliography}{10}

\bibitem{Achucarro-T} A. Ach\'ucarro and 
P.K. Townsend,  Phys. Lett. {\bf B180}, 89 (1986).

\bibitem{Witten88} E. Witten,  Nucl. Phys. {\bf B 311}, 4
(1988).

\bibitem{BH} J.D. Brown and M. Henneaux, Commun.
Math. Phys. {\bf 104}, 207 (1986).

\bibitem{CHvD}
O. Coussaert, M. Henneaux, P. van Driel 
 Class. Quant. Grav.  {\bf 12}, 2961 (1995). 

\bibitem{Banados95}   M. Ba\~nados,  Phys. Rev. {\bf D52}, 5816 (1995).

\bibitem{Carlip} S. Carlip, Phys. Rev. {\bf D51}, 632 (1995)
; S. Carlip, Phys. Rev. {\bf D55}, 878 (1997)

\bibitem{BBO} M. Ba\~nados, T. Brotz and M. Ortiz, 
Nucl.Phys.{\bf B545}, 3470 (1999)
[hep-th/9802076] 

\bibitem{Mansouri-}
S.~Fernando and F.~Mansouri,
Phys.\ Lett.\ B {\bf 505}, 206 (2001)
[hep-th/0010153].

\bibitem{Chamseddine} A.H. Chamseddine, 
Nucl. Phys. {\bf B346}, 213 (1990).

\bibitem{TZ} R. Troncoso and J. Zanelli, 
Phys.Rev. {\bf D58}, R101703 (1998);
R.~Troncoso and J.~Zanelli, hep-th/9902003.

\bibitem{BTrZ} M. Ba\~nados, R. Troncoso and J. 
        Zanelli,  Phys. Rev. {\bf D54}, 2605 (1996).  


\bibitem{Nair-}  V.P. Nair and J. Schiff, Phys. Lett. {\b B246}, 
     423, (1990);  Nucl. Phys. {\bf B371}, 329 (1992).   

\bibitem{BGH} M. Ba\~nados, L.J. Garay and M. Henneaux, 
              Phys. Rev. {\bf D53}, R593 (1996); 
              M. Ba\~nados, L.J. Garay and M. Henneaux,
              Nucl. Phys. {\bf B476}, 611 (1996).

\bibitem{Losev}
A.~Losev, G.~Moore, N.~Nekrasov and S.~Shatashvili,
Nucl.\ Phys.\ Proc.\ Suppl.\  {\bf 46}, 130 (1996); 
 Nucl.Phys.B484:196-222,1997 


\bibitem{Gegenberg-K}
J.~Gegenberg and G.~Kunstatter,
Phys.\ Lett.\  {\bf B478}, 327 (2000)

\bibitem{11d} M. Ba\~nados, hep-th/0107214 

\bibitem{Horava}
P.~Horava,
Phys.\ Rev.\ {\bf D59}, 046004 (1999)
[hep-th/9712130].

\bibitem{CJS} E. Cremmer, B. Julia and J. Scherk 
 Phys.Lett. {\bf B76}, 409 (1978)

\bibitem{Nahm}
W.~Nahm,
Nucl.\ Phys.\  {\bf B135}, 149 (1978).

\bibitem{BTZ2} M. Ba\~nados, C. Teitelboim and J.Zanelli,
               Phys. Rev. {\bf D49}, 975 (1994).

\bibitem{dAuria}
L.~Castellani, R.~D' Auria and P.~Fre,
``Supergravity and superstrings: 
A Geometric perspective. Vol. 2: Supergravity,''
{\it  Singapore, Singapore: World Scientific (1991) 607-1371}.

\bibitem{Sardinia} 
M.~Ba\~nados,
Nucl.\ Phys.\ Proc.\ Suppl.\  {\bf 88}, 17 (2000)
[hep-th/9911150].

\bibitem{Boulware-D}
D.G.~Boulware and S.~Deser,
Phys.\ Rev.\ Lett.\ {\bf 55}, 2656 (1985).

\bibitem{Wheeler} J.T. Wheeler, {\em Nucl.Phys.} {\bf B268}, 737
     (1986); {\bf B273}, 732 (1986). 

\bibitem{BTZ} M. Ba\~nados, C. Teitelboim and J.Zanelli,
              Phys. Rev. Lett. {\bf 69}, 1849 (1992).  
              Ba\~nados, M. Henneaux, C. Teitelboim and
              J.Zanelli,  Phys. Rev. {\bf D48}, 1506 (1993).

\bibitem{selfdual}
M.~Ba\~nados, 
Phys.\ Lett.\  {\bf B501}, 150 (2001)
[hep-th/0009066].


\end{thebibliography}
\end{document}